\documentclass{article}

\usepackage{arxiv}

\usepackage[utf8]{inputenc} 
\usepackage[T1]{fontenc}    
\usepackage{hyperref}       
\usepackage{url}            
\usepackage{booktabs}       
\usepackage{amsfonts}       
\usepackage{nicefrac}       
\usepackage{microtype}      
\usepackage{lipsum}		
\usepackage{graphicx}
\usepackage{epstopdf}
\usepackage{subfigure}
\usepackage{amsmath}

\newtheorem{theorem}{Theorem}[section]

\newtheorem{remark}[theorem]{Remark}

\title{Boundary-free Kernel-smoothed Goodness-of-fit Tests for Data on General Interval}

\author{ \href{https://orcid.org/0000-0002-6939-9465}{\includegraphics[scale=0.06]{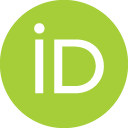}\hspace{1mm}Rizky Reza Fauzi}\thanks{corresponding author}\\
	Graduate School of Mathematics\\
	Kyushu University\\
	744 Motooka, Nishi-ku, Fukuoka-shi, Fukuoka-ken, JAPAN\\
	\texttt{fauzi.rizky.853@s.kyushu-u.ac.jp}\\
	\And
	Yoshihiko Maesono\\
	Faculty of Sciemce and Engineering\\
	Chuo University\\
	1-13-27 Kasuga, Bunkyo-ku, Tokyo-to, JAPAN\\
	\texttt{maesono@math.chuo-u.ac.jp}\\
}

\hypersetup{
pdftitle={A template for the arxiv style},
pdfsubject={q-bio.NC, q-bio.QM},
pdfauthor={David S.~Hippocampus, Elias D.~Striatum},
pdfkeywords={First keyword, Second keyword, More},
}

\begin{document}
\maketitle

\begin{abstract}
We propose kernel-type smoothed Kolmogorov-Smirnov and Cram\'{e}r-von Mises tests for data on general interval, using bijective transformations. Though not as severe as in the kernel density estimation, utilizing naive kernel method directly to those particular tests will result in boundary problem as well. This happens mostly because the value of the naive kernel distribution function estimator is still larger than $0$ (or less than $1$) when it is evaluated at the boundary points. This situation can increase the errors of the tests especially the second-type error. In this article, we use bijective transformations to eliminate the boundary problem. Some simulation results illustrating the estimator and the tests' performances will be presented in the last part of this article.
\end{abstract}


\section{Introduction}
Many statistical methods depend on an assumption that the data under consideration are drawn from a certain distribution, or at least from a distribution that is approximately similar to that particular distribution. For example, test of normality for residuals are needed after fitting a linear regression in order to satisfy the normality assumption of the model. Distributional assumption is important because, in most cases, it dictates the methods that can be used to estimate the unknown parameters and also determines the procedures that staticticians may apply. There are some goodness-of-fit tests available to determine whether a sample comes from the assumed distribution. Those popular tests include the Kolmogorov-Smirnov (KS) test, Cram\'{e}r-von Mises (CvM) test, Anderson-Darling test, and Durbin-Watson test. In this article, we will be focusing ourselves to the KS and CvM tests.

Let $X_1,X_2,...,X_n$ be independently and identically distributed random variables supported on $\Omega\subseteq\mathbb{R}$ with an absolutely continuous distribution function $F_X$ and a density $f_X$. The classical nonparametric estimator of $F_X$ has been the empirical distribution function defined by
\begin{gather}
F_n(x)=\frac{1}{n}\sum_{i=1}^n I(X_i\leq x), \; \; \; x\in\mathbb{R},
\end{gather}
where $I(A)$ denotes the indicator function of a set $A$. It is obvious that $F_n$ is a step function of height $\frac{1}{n}$ at each observed sample point $X_i=x_i$. When considered as a pointwise estimator of $F_X(x)$, $F_n(x)$ is an unbiased and strongly consistent estimator with $Var[F_n(x)]=n^{-1}F_X(x)[1-F_X(x)]$.

In this setting, the Kolmogorov-Smirnov statistic utilizes the empirical distribution function $F_n$ to test the null hypothesis
\begin{gather*}
H_0: F_X=F
\end{gather*}
againsts the alternative hypothesis
\begin{gather*}
H_1: F_X\neq F,
\end{gather*}
where $F$ is the assumed distribution function. The test statistic is defined as
\begin{gather}
KS_n=\sup_{x\in\mathbb{R}}|F_n(x)-F(x)|.
\end{gather}
If under a significance level $\alpha$ the value of $KS_n$ is larger than a certain value from Kolmogorov distribution table, we will reject $H_0$. Likewise, under the same circumstance, the statistic of the Cram\'{e}r-von Mises test is defined as
\begin{gather}
CvM_n=n\int_{-\infty}^\infty [F_n(x)-F(x)]^2\mathrm{d}F(x),
\end{gather}
and we reject the null hypothesis when the value of $CvM_n$ is larger than a certain value from Cram\'{e}r-von Mises table.

Several discussions regarding those goodness-of-fit tests have been around for decades. The recent articles include the distribution of KS and CvM tests for exponential populations (Evans \textit{et al}. 2017), revision of two-sample KS test (Finner and Gontscharuk 2018), KS test for mixed distributions (Zierk \textit{et al}. 2020), KS test for bayesian ensembles of phylogenies (Antoneli \textit{et al}. 2018), CvM distance for neighbourhood-of-model validation (Baringhaus and Henze 2016), rank-based CvM test (Curry \textit{et al}. 2019), and model selection using CvM distance in a fixed design regression (Chen \textit{et al}. 2018).

Though the standard KS and CvM tests work really well, but it does not mean they bear no problem. The lack of smoothness of $F_n$ causes too much sensitivity near the center of distribution, especially when $n$ is small. Hence, it is not unusual to find the supremum value of $|F_n(x)-F(x)|$ is attained when $x$ is near the center of distribution, or the value of $CvM_n$ gets larger because $[F_n(x)-F(x)]^2$ is large when the data is highly concentrated in one area. Furthermore, given the information that $F_X$ is absolutely continuous, it seems to be more appropriate to use a smooth and continuous estimator of $F_X$ rather than the empirical distribution function $F_n$ for testing the goodness-of-fit.

The other maneuver that can be used for estimating the cummulative distribution function nonparametrically is the kernel method. Let $K(x)$ be a symmetric continuous nonnegative kernel function with $\int_{-\infty}^\infty K(x)\mathrm{d}x=1$, and $h>0$ be the bandwidth satisfying $h\rightarrow 0$ and $nh\rightarrow\infty$ when $n\rightarrow\infty$. Hence, Nadaraya (1964) defined the naive kernel distribution function estimator as
\begin{gather}
\widehat{F}_X(x)=\frac{1}{n}\sum_{i=1}^n W\left(\frac{x-X_i}{h}\right), \; \; \; x\in\mathbb{R},
\end{gather}
where $W(v)=\int_{-\infty}^v K(w)\mathrm{d}w$. It is easy to prove that this kernel distribution function estimator is continuous, and satisfies all the properties of a distribution function.

Several properties of $\widehat{F}_X$ are well known. The bias and the variance are
\begin{gather}
Bias[\widehat{F}_X(x)]=\frac{h^2}{2}f_X'(x)\int_{-\infty}^\infty v^2 K(v)\mathrm{d}v+o(h^2),\\
Var[\widehat{F}_X(x)]=\frac{1}{n}F_X(x)[1-F_X(x)]-\frac{2h}{n}r_1 f_X(x)+o\left(\frac{h}{n}\right),
\end{gather}
where $r_1=\int_{-\infty}^\infty vK(v)W(v)\mathrm{d}v$. The almost sure uniform convergence of $\widehat{F}_X$ to $F_X$ was proved by Nadaraya (1964), Winter (1973), and Yamato (1973), while Yukich (1989) extended this result to higher dimensions. Watson and Leadbetter (1964) proved the asymptotic normality of $\widehat{F}_X(x)$. Moreover, several authors showed that the asymptotic performance of $\widehat{F}_X(x)$ is better than that of $F_n(x)$, see Azzalini (1981), Reiss (1981), Falk (1983), Singh \textit{et al}. (1983), Hill (1985), Swanepoel (1988), Shirahata and Chu (1992), and Abdous (1993).

It is natural if one uses the naive kernel distribution function estimator in place of the empirical distribution function to smooth the KS and CvM statistics out. By doing that, we may expect to eliminate the over-sensitivity that standard KS and CvM statistics have. Therefore, the formulas become
\begin{gather}
\widehat{KS}=\sup_{x\in\mathbb{R}}|\widehat{F}_X(x)-F(x)|
\end{gather}
and
\begin{gather}
\widehat{CvM}=n\int_{-\infty}^\infty [\widehat{F}(x)-F(x)]^2\mathrm{d}F(x).
\end{gather}
Omelka \textit{et al}. (2009) proved that under the null hypothesis, the distribution of the statistics converge to the same distributions as the standard KS and CvM.

Though both tests are versatile in most settings, but when the support of the data is strictly smaller than the entire real line, the naive kernel distribution function estimator suffers the so called boundary problem. This problem happens because the estimator still puts some weights outside the support $\Omega$. Even though in some cases (e.g. $f_X(0)=0$ when $0$ is the boundary point) the boundary effects of $\widehat{F}_X(x)$ is not as severe as in the kernel density estimator, but the problem still occurs. It is because the value of $\widehat{F}_X(x)$ is still larger than $0$ (or less than $1$) at the boundary points. This phenomena cause large value of $|\widehat{F}_X(x)-F(x)|$ in the boundary regions, and then $\widehat{KS}$ and $\widehat{CvM}$ tend to be larger than they are supposed to be, leading to the rejection of $H_0$ even though $H_0$ is right. To make things worse, section 4 will illustrate how this problem enlarges type-2 error by accepting the null hypothesis when it is wrong.

Some articles have suggested methods to solve the boundary bias problem in the density estimation, such as data reflection by Schuster (1985); simple nonnegative boundary correction by Jones and Foster (1996); boundary kernels by M\"{u}ller (1991), M\"{u}ller (1993), and M\"{u}ller and Wang (1994); generating pseudo data by Cowling and Hall (1996); and hybrid method by Hall and Wehrly (1991). Even though only few literature discusses how to extend previous ideas for solving the problem in the distribution function estimation, but it is reasonable to assume those methods are applicable for such case.

In this article we will try another idea to remove the boundary effects, which is utilizing bijective mappings. How we remove the boundary effects from the naive kernel distribution function estimator will be discussed in section 2, and how to modify the goodness-of-fit tests with those idea will be explained in section 3. Some numerical studies are discussed in section 4, and the proofs of our theorems can be found in the appendices.

\section{Boundary-free kernel distribution function estimator}
In this section, we will explain how to use bijective transformations to solve the boundary problem in kernel distribution function estimation. It is obvious that if we can find an appropiate function $g$ that maps $\mathbb{R}$ to $\Omega$ bijectively, we will not put any weight outside the support. Hence, instead of using $X_1,X_2,...,X_n$, we will apply the kernel method for $g^{-1}(X_1),g^{-1}(X_2),...,g^{-1}(X_n)$. To make sure our idea is mathematically applicable, we need to impose some conditions before moving on to our main focus. The conditions we took are:
\begin{enumerate}
	\item[\textbf{C1}.] the kernel function $K(v)$ is nonnegative, continuous, and symmetric at $v=0$
	\item[\textbf{C2}.] the integral $\int_{-\infty}^\infty v^2 K(v)\mathrm{d}v$ is finite and $\int_{-\infty}^\infty K(v)\mathrm{d}v=1$
	\item[\textbf{C3}.] the bandwidth $h>0$ satisfies $h\rightarrow 0$ and $nh\rightarrow\infty$ when $n\rightarrow\infty$
	\item[\textbf{C4}.] the increasing function $g$ transforms $\mathbb{R}$ onto $\Omega$
	\item[\textbf{C5}.] the density $f_X$ and the function $g$ are twice differentiable
\end{enumerate}
The conditions C1-C3 are standard conditions for kernel method. Albeit it is sufficient for $g$ to be a bijective function, but the increasing property in C4 makes the proofs of our theorems simpler. The last condition is needed to derive the biases and the variances formula.

Under those conditions, we define the boundary-free kernel distribution function estimator as
\begin{gather}
\widetilde{F}_X(x)=\frac{1}{n}\sum_{i=1}^n W\left(\frac{g^{-1}(x)-g^{-1}(X_i)}{h}\right), \; \; \; x\in\Omega
\end{gather}
where $h>0$ is the bandwidth and $g$ is an appropriate bijective function. As we can see, $\widetilde{F}_X(x)$ is basically just a result of simple subsitution of $g^{-1}(x)$ and $g^{-1}(X_i)$ to the formula of $\widehat{F}_X(x)$. Though it looks simple, but the argument behind this idea is due to the change-of-variable property of distribution function, which cannot always be done to another probability-related functions. Its bias and variance are given in the following theorem.
\begin{theorem}\label{thm:biasvar}
	Under the conditions C1-C5, the bias and the variance of $\widetilde{F}_X(x)$ are
	\begin{gather}
	Bias[\widetilde{F}_X(x)]=\frac{h^2}{2}c_1(x)\int_{-\infty}^\infty v^2 K(v)\mathrm{d}v+o(h^2),\\
	Var[\widetilde{F}_X(x)]=\frac{1}{n}F_X(x)[1-F_X(x)]-\frac{2h}{n}g'(g^{-1}(x))f_X(x)r_1+o\left(\frac{h}{n}\right),
	\end{gather}
	where
	\begin{gather}
	c_1(x)=g''(g^{-1}(x))f_X(x)+[g'(g^{-1}(x))]^2 f_X'(x).
	\end{gather}
\end{theorem}
\begin{remark}\label{rem:biasvar}
	It is easy to prove that $r_1$ is a positive number. Then, since $g$ is an increasing function, the variance of our proposed estimator will be smaller than $Var[\widehat{F}_X(x)]$ when $g'(g^{-1}(x))\geq 1$. On the other hand, though it is difficult to conclude in general case, if we carefully take the mapping $g$, the bias of our proposed method is much faster to converge to $0$ than $Bias[\widehat{F}_X(x)]$. For example, when $\Omega=\mathbb{R}^+$ and we choose $g(x)=e^x$, in the boundary region when $x\rightarrow 0$ the bias will converge to $0$ faster and $Var[\widetilde{F}_X(x)]<Var[\widehat{F}_X(x)]$.
\end{remark}

Similar to most of kernel type estimators, our proposed estimator attains asymptotic normality, as stated in the following theorem.
\begin{theorem}\label{thm:normal}
	Under the condition C1-C5, the limiting distribution
	\begin{gather}
	\frac{\widetilde{F}_X(x)-F_X(x)}{\sqrt{Var[\widetilde{F}_X(x)]}}\rightarrow_D N(0,1)
	\end{gather}
	holds.
\end{theorem}
Furthermore, we also establish strong consistency of the proposed method.
\begin{theorem}\label{thm:konsisten}
	Under the condition C1-C5, the consistency
	\begin{gather}
	\sup_{x\in\Omega}|\widetilde{F}_X(x)-F_X(x)|\rightarrow_{a.s.}0
	\end{gather}
	holds.
\end{theorem}

Even though it is not exactly related to our main topic of goodness-of-fit tests, but it is worth to add that from $\widetilde{F}_X$ we can derive another kernel-type estimator. It is clear that the density function $f_X$ is equal to $F_X'$, then we can define a boundary-free kernel density estimator as $\widetilde{f}_X=\frac{\mathrm{d}}{\mathrm{d}x}\widetilde{F}_X$, which is
\begin{gather}
\widetilde{f}_X(x)=\frac{1}{nhg'(g^{-1}(x))}\sum_{i=1}^n K\left(\frac{g^{-1}(x)-g^{-1}(X_i)}{h}\right), \; \; \; x\in\Omega.
\end{gather}
As $\widetilde{F}_X$ eliminates boundary bias problem, this new estimator $\widetilde{f}_X$ does the same thing and can be a good competitor for another boundary bias reduction kernel density estimators. The bias and the variance of its are as follow.
\begin{theorem}\label{thm:biasvarpdf}
	Under the condition C1-C5, also if $g'''$ exists and $f_X''$ is continuous, then the bias and the variance of $\widetilde{f}_X(x)$ are
	\begin{gather}
	Bias[\widetilde{f}_X(x)]=\frac{h^2 c_2(x)}{2g'(g^{-1}(x))}\int_{-\infty}^\infty v^2 K(v)\mathrm{d}v+o(h^2)\\
	Var[\widetilde{f}_X(x)]=\frac{f_X(x)}{nhg'(g^{-1}(x))}\int_{-\infty}^\infty K^2(v)\mathrm{d}v+o\left(\frac{1}{nh}\right),
	\end{gather}
	where
	\begin{gather}
	c_2(x)=g'''(g^{-1}(x))f_X(x)+3g''(g^{-1}(x))g'(g^{-1}(x))f_X'(x)+[g'(g^{-1}(x))]^3 f_X''(x).
	\end{gather}
\end{theorem}

\section{Boundary-free kernel-smoothed KS and CvM tests}
As we discussed before, the problem of the standard KS and CvM statistics is in the over-sensitivity near the center of distribution, because of the lack of smoothness of the empirical distribution function. Since the area around the center of distribution has the highest probability density, most of the realizations of the sample are there. As a result, $F_n(x)$ jumps a lot in those area, and it causes some unstability of estimation especially when $n$ is small. Conversely, though smoothing $KS_n$ and $CvM_n$ out using kernel distribution function can eliminate the oversensitivity near the center, the value of $\widehat{KS}$ and $\widehat{CvM}$ become larger than it should be when the data we are dealing with is supported on an interval smaller than the entire real line. This phenomenon is caused by the boundary problem.

Therefore, the clear solution to overcome the problems of standard and naive kernel goodness-of-fit tests together is to keep the smoothness of $\widehat{F}_X$ and to get rid of the boundary problem simulateously. One of the idea is by utilizing the boundary-free kernel distribution function estimator in section 2. Therefore, we propose boundary-free kernel-smoothed Kolmogorov-Smirnov statistic as
\begin{gather}
\widetilde{KS}=\sup_{x\in\mathbb{R}}|\widetilde{F}_X(x)-F(x)|
\end{gather}
and boundary-free kernel-smoothed Cram\'{e}r-von Mises statistic as
\begin{gather}
\widetilde{CvM}=n\int_{-\infty}^\infty[\widetilde{F}_X(x)-F(x)]^2\mathrm{d}F(x),
\end{gather}
where $\widetilde{F}_X$ is our proposed estimator with a suitable function $g$.
\begin{remark}
	Although the supremum and the integral are evaluated througout the entire real line, but we can just compute them over $\Omega$, as on the outside of the support we have $F_X(x)=\widetilde{F}_X(x)$.
\end{remark}

Although the formulas seem similar, one might expect both proposed tests are totally different with the standard KS and CvM tests. However, these two following theorems explain that the standard ones and our proposed methods turn out to be equivalent in the sense of distribution.
\begin{theorem}\label{thm:ksequivalent}
	Let $F_X$ and $F$ be distribution functions on set $\Omega$. If $KS_n$ and $\widetilde{KS}$ are the standard and the proposed Kolmogorov-Smirnov statistics, respectively, then under the null hypothesis $F_X=F$,
	\begin{gather}
	|KS_n-\widetilde{KS}|\rightarrow_p 0.
	\end{gather}
\end{theorem}
\begin{theorem}\label{thm:cvmequivalnent}
	Let $F_X$ and $F$ be distribution functions on set $\Omega$. If $CvM_n$ and $\widetilde{CvM}$ are the standard and the proposed Cram\'{e}r-von Mises statistics, respectively, then under the null hypothesis $F_X=F$,
	\begin{gather}
	|CvM_n-\widetilde{CvM}|\rightarrow_p 0.
	\end{gather}
\end{theorem}
Those equivalencies allow us to use the same distribution tables of the standard goodness-of-fit tests for our new statistics. It means, with the same significance level $\alpha$, the critical values are same.

\section{Numerical studies}
We will show the results of our numerical studies in this section. The studies consist of two parts, the simulations of the proposed estimators $\widetilde{F}_X$ and $\widetilde{f}_X$, and then the results of the new goodness-of-fit tests $\widetilde{KS}$ and $\widetilde{CvM}$.

\subsection{boundary-free kernel DF and PDF estimations results}
For the simulation to show the performances of the new distribution function estimator, we calculated the average integrated squared error (AISE) and repeated them $1000$ times for each case. We compared the naive kernel distribution function estimator $\widehat{F}_X$ and our proposed estimator $\widetilde{F}_X$. In the case of the proposed method, we chose two mappings $g^{-1}$ for each case. When $\Omega=\mathbb{R}^+$, we used the logarithm function $\log(x)$ and a composite of two functions $\Phi^{-1}\circ\gamma$, where $\gamma(x)=1-e^x$. However, if $\Omega=[0,1]$, we utilized probit and logit functions. With size $50$, the generated samples were drawn from gamma $Gamma(2,2)$, weibull $Weibull(2,2)$, standard log-normal $\log.N(0,1)$, absolute-normal $abs.N(0,1)$, standard uniform $U(0,1)$, and beta distributions with three different sets of parameters ($Beta(1,3)$, $Beta(2,2)$, and $Beta(3,1)$). The kernel function we used here is the Gaussian Kernel and the bandwidths were chosen by cross-validation technique. We actually did the same simulation study using the Epanechnikov Kernel, but the results are quite similar. Graphs of some chosen cases are shown as well in Figure 1.

\begin{table}
	\centering
	\caption{AISE ($\times 10^5$) comparison of DF estimators}
	\label{tab:aise_df}
	\begin{tabular}{llllll}
		\hline\noalign{\smallskip}
		Distributions&$\widehat{F}_X$&$\widetilde{F}_{\log}$&$\widetilde{F}_{\Phi^{-1}\circ\gamma}$&$\widetilde{F}_{probit}$&$\widetilde{F}_{logit}$\\
		\noalign{\smallskip}\hline\noalign{\smallskip}
		$Gamma(2,2)$&$2469$&$2253$&$\mathbf{2181}$&-&-\\
		$Weibull(2,2)$&$2224$&$\mathbf{1003}$&$1350$&-&-\\
		$\log.N(0,1)$&$1784$&$1264$&$\mathbf{1254}$&-&-\\
		$abs.N(0,1)$&$2517$&$\mathbf{544}$&$727$&-&-\\
		$U(0,1)$&$5074$&-&-&$\mathbf{246}$&$248$\\
		$Beta(1,3)$&$7810$&-&-&$\mathbf{170}$&$172$\\
		$Beta(2,2)$&$6746$&-&-&$\mathbf{185}$&$188$\\
		$Beta(3,1)$&$7801$&-&-&$\mathbf{154}$&$156$\\
		\noalign{\smallskip}\hline
	\end{tabular}
\end{table}

\begin{figure}
	\centering
	\subfigure[$X\sim Gamma(2,2)$\label{fig:dfgamma}]{
		\resizebox*{5cm}{!}{\includegraphics{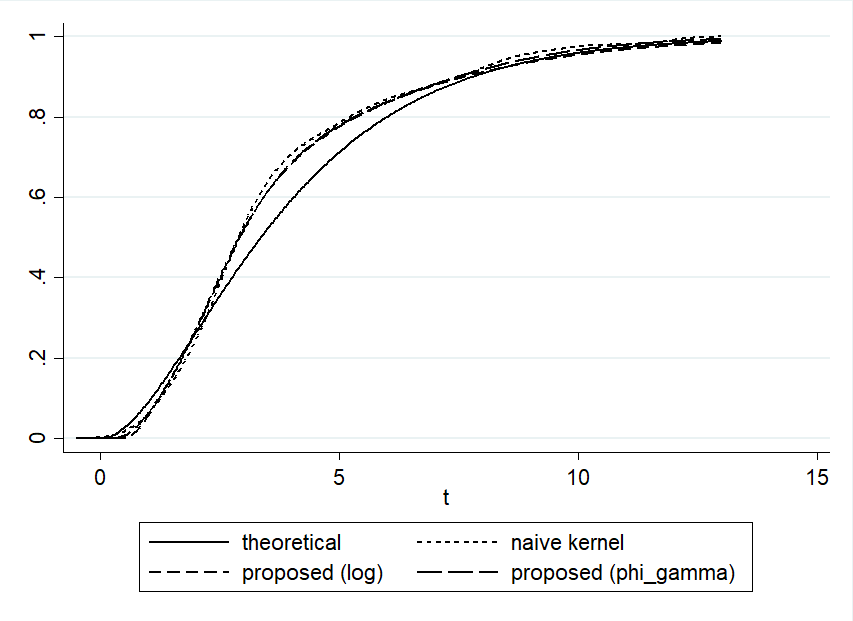}}}
	\subfigure[$X\sim abs.N(0,1)$\label{fig:dfabsn}]{
		\resizebox*{5cm}{!}{\includegraphics{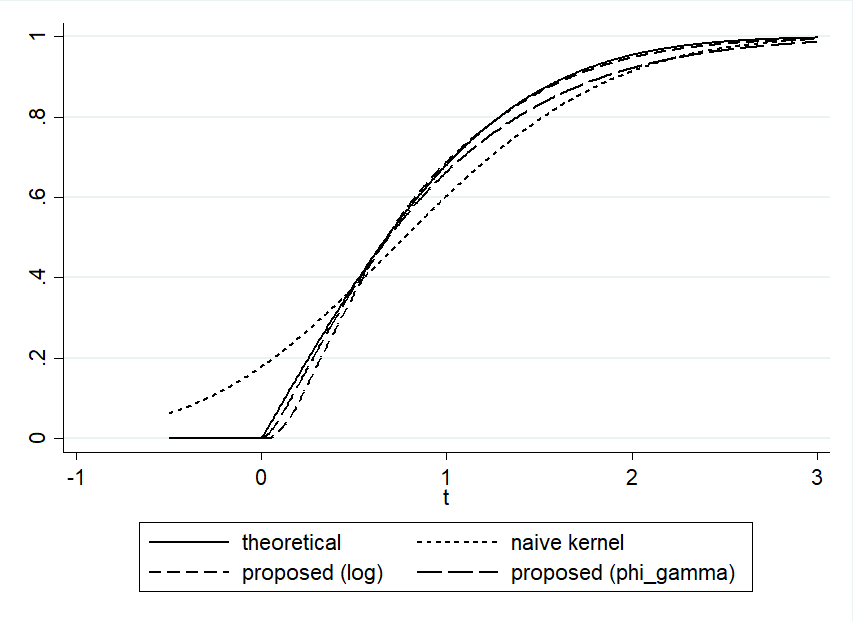}}}\\
	\subfigure[$X\sim U(0,1)$\label{fig:dfu}]{
		\resizebox*{5cm}{!}{\includegraphics{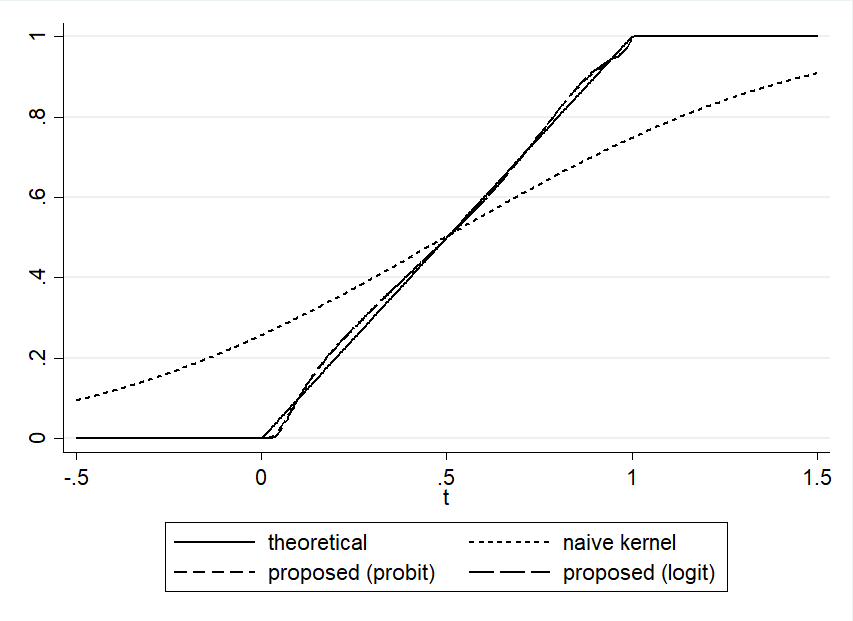}}}
	\subfigure[$X\sim Beta(2,2)$\label{fig:dfbeta}]{
		\resizebox*{5cm}{!}{\includegraphics{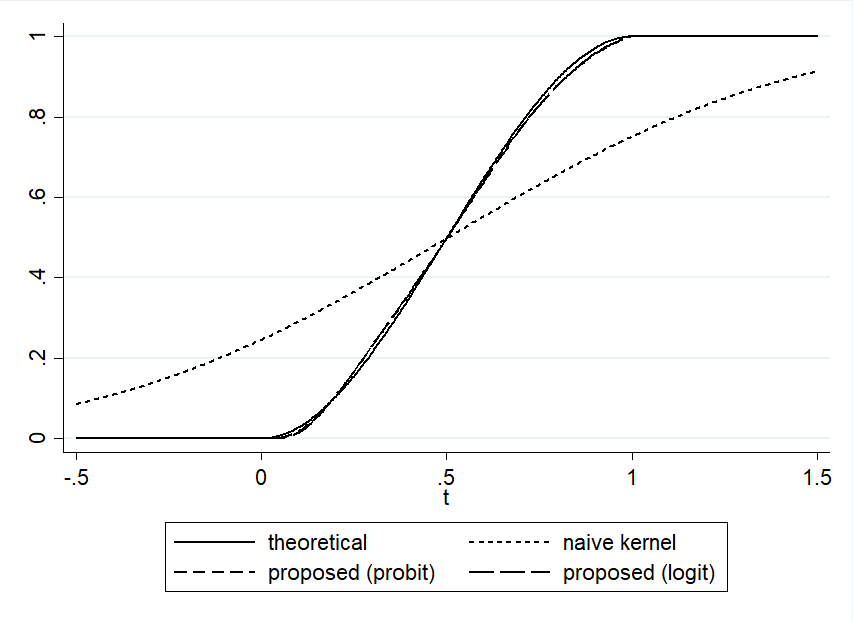}}}
	\caption{Graphs comparisons of $F_X(x)$, $\widehat{F}_X(x)$, and $\widetilde{F}_X(x)$ for several distributions, with sample size $n=50$.}\label{fig:df}
\end{figure}

As we can see in Table 1, our proposed estimator outperformed the naive kernel distribution function. Though the differences are not so big in the cases of gamma, weibull, and the log-normal distributions, but the gaps are glaring in the absolute-normal case or when the support of the distributions is the unit interval. The cause of this phenomena might be seen in Figure 1.

Albeit the shapes of $\widetilde{F}_{\log}$ and $\widetilde{F}_{\Phi^{-1}\circ\gamma}$ are more similar to the theoretical distribution in Figure 1(a), but we have to admit that the shape of $\widehat{F}_X$ is not so much different with the rests. However in Figure 1(b), (c), and (d), it is obvious that the naive kernel distribution function is too far-off the mark, particularly in the case of $\Omega=[0,1]$. As the absolute-normal, uniform, and beta distributions have quite high probability density near the boundary point $x=0$ (also $x=1$ for unit interval case), the naive kernel estimator spreads this "high density information" around the boundary regions. However, since $\widehat{F}_X$ cannot detect the boundaries, it puts this "high density information" outside the support as well. This is not happening too severely in the case of Figure 1(a) because the probability density near $x=0$ is fairly low. Hence, although the value of $\widehat{F}_X(x)$ might be still positive when $x\approx 0^-$, but it is not so far from $0$ and vanishes quickly
\begin{remark}
	Figure 1(c) and (d) also gave a red-alert if we try to use the naive kernel distribution function in place of empirical distribution for goodness-of-fit tests. As the shapes of $\widehat{F}_X$ in Figure 1(c) and (d) resemble the normal distribution function a lot, if we test $H_0: X\sim N(\mu,\sigma^2)$, we will find the tests may not reject the null hypothesis. This shall cause the increment of type-2 error.
\end{remark}
\begin{remark}
	It is worth to note that in Table 1, even though $\widetilde{F}_{probit}$ performed better, but its differences are too little to claim that it outperformed $\widetilde{F}_{logit}$. From here we can conclude that probit and logit functions work pretty much the same for $\widetilde{F}_X$.
\end{remark}

Since we introduced $\widehat{f}_X$ as a new boundary-free kernel density estimator, we also provide some illustrations of its performances in this subsection. Under the same settings as in the simulation study of the distribution function case, we can see the results of its simulation in Table 2 and Figure 2.

\begin{table}
	\centering
	\caption{AISE ($\times 10^5$) comparison of density estimators}
	\label{tab:aise_pdf}
	\begin{tabular}{llllll}
		\hline\noalign{\smallskip}
		Distributions&$\widehat{f}_X$&$\widetilde{f}_{\log}$&$\widetilde{f}_{\Phi^{-1}\circ\gamma}$&$\widetilde{f}_{probit}$&$\widetilde{f}_{logit}$\\
		\noalign{\smallskip}\hline\noalign{\smallskip}
		$Gamma(2,2)$&$925$&$744$&$\mathbf{624}$&-&-\\
		$Weibull(2,2)$&$6616$&$\mathbf{3799}$&$3986$&-&-\\
		$\log.N(0,1)$&$7416$&$3569$&$\mathbf{2638}$&-&-\\
		$abs.N(0,1)$&$48005$&$34496$&$\mathbf{14563}$&-&-\\
		$U(0,1)$&$36945$&-&-&$\mathbf{14235}$&$21325$\\
		$Beta(1,3)$&$109991$&-&-&$\mathbf{18199}$&$28179$\\
		$Beta(2,2)$&$52525$&-&-&$\mathbf{5514}$&$6052$\\
		$Beta(3,1)$&$109999$&-&-&$\mathbf{17353}$&$28935$\\
		\noalign{\smallskip}\hline
	\end{tabular}
\end{table}

\begin{figure}
	\centering
	\subfigure[$X\sim Gamma(2,2)$\label{fig:pdfgamma}]{
		\resizebox*{5cm}{!}{\includegraphics{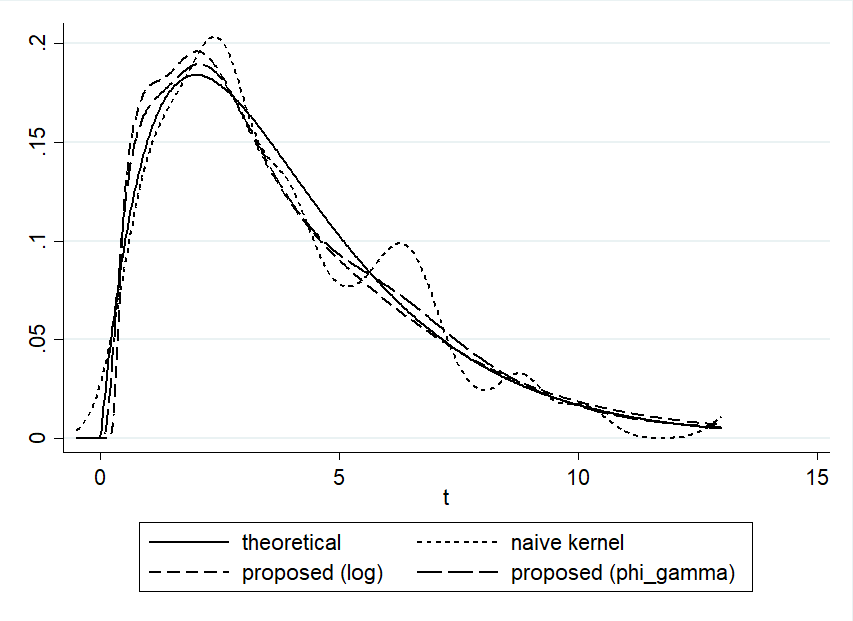}}}
	\subfigure[$X\sim abs.N(0,1)$\label{fig:pdfabsn}]{
		\resizebox*{5cm}{!}{\includegraphics{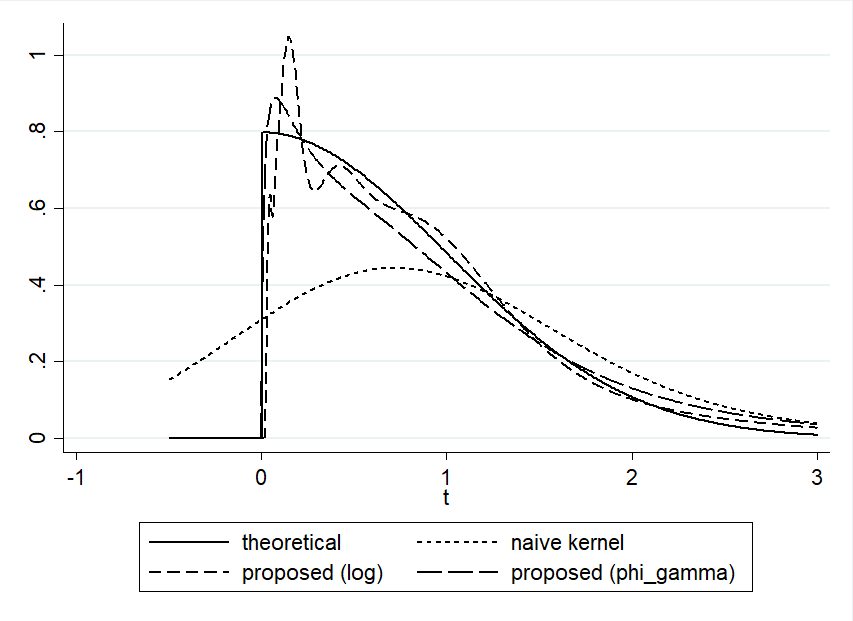}}}\\
	\subfigure[$X\sim U(0,1)$\label{fig:pdfu}]{
		\resizebox*{5cm}{!}{\includegraphics{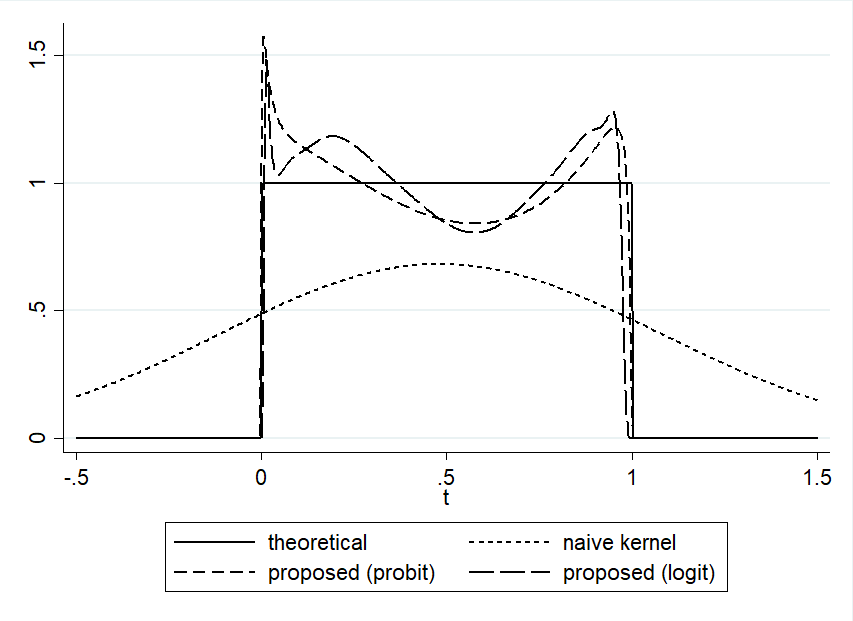}}}
	\subfigure[$X\sim Beta(2,2)$\label{fig:pdfbeta}]{
		\resizebox*{5cm}{!}{\includegraphics{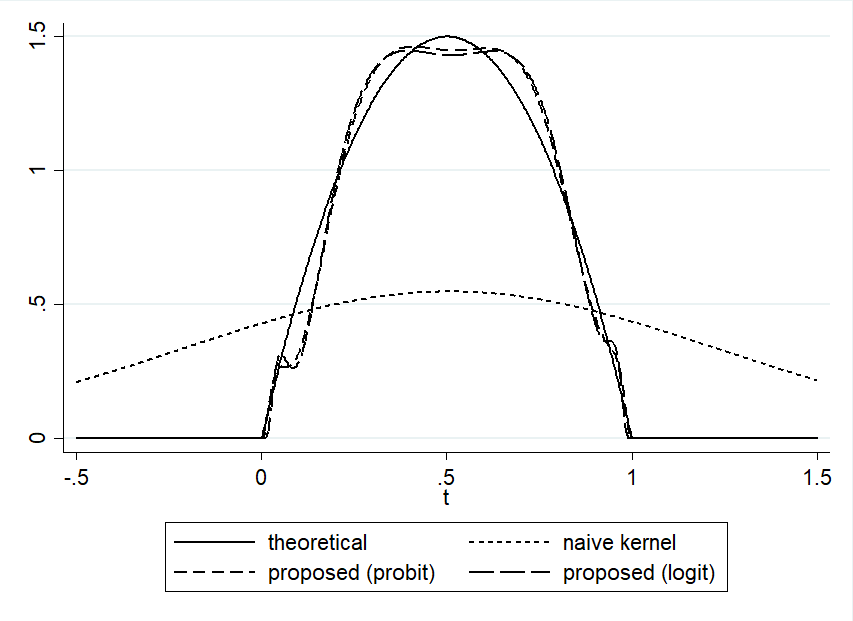}}}
	\caption{Graphs comparisons of $f_X(x)$, $\widehat{f}_X(x)$, and $\widetilde{f}_X(x)$ for several distributions, with sample size $n=50$.}\label{fig:pdf}
\end{figure}

From AISE point of view, once again our proposed estimator outperformed the naive kernel one, and huge gaps happened as well when the support of the distribution is the unit interval. We may take some interests in Figure 2(b), (c), and (d), as the graphs of $\widehat{F}_X$ are too different with the theoretical ones, and more similar to the gaussian bell shapes instead. This situation resonates with our claim in Remark 4.1.

\subsection{boundary-free kernel-type KS and CvM tests simulations}
We provide the results of our simulation studies regarding the new Kolmogorov-Smirnov and Cram\'{e}r-von Mises tests in this part. As a measure of comparison, we calculated the percentage of rejecting several null hypothesis when the samples were drawn from certain distributions. When the actual distribution and the null hypothesis are same, we expect the percentage should be close to $100\alpha\%$ (significance level in percent). However, if the real distribution does not match the $H_0$, we hope to see the percentage is as large as possible. To illustrate how the behaviours of the statistics change, we generated a sequential numbers of sample sizes, started from $10$ until $100$, with $1000$ repetitions for each case. We chose level of significance $\alpha=0.01$, and we compared the standard KS and CvM tests with our proposed tests.

\begin{figure}
	\centering
	\subfigure[$H_0:Gamma(2,2)$\label{fig:weibull vs gamma}]{
		\resizebox*{5cm}{!}{\includegraphics{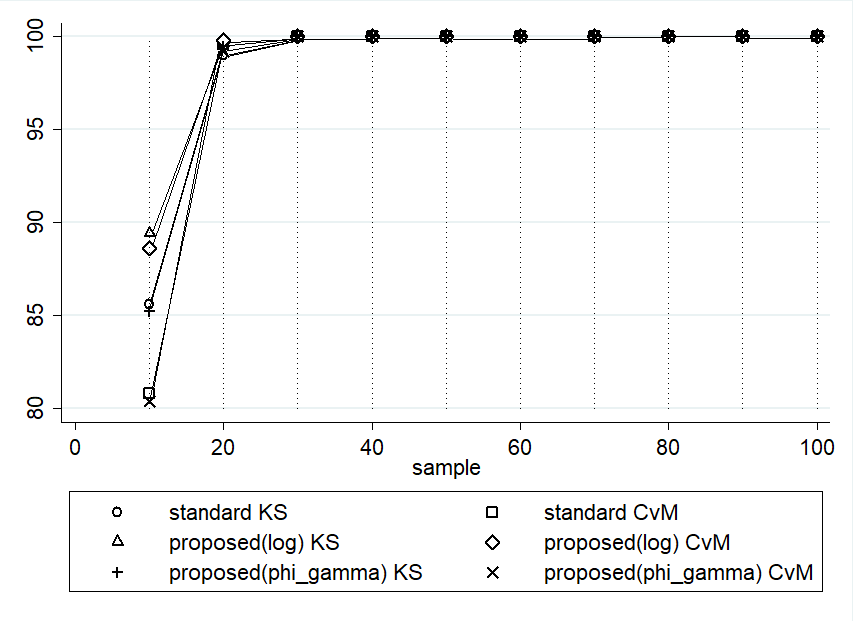}}}
	\subfigure[$H_0:Weibull(2,2)$\label{fig:weibull vs weibull}]{
		\resizebox*{5cm}{!}{\includegraphics{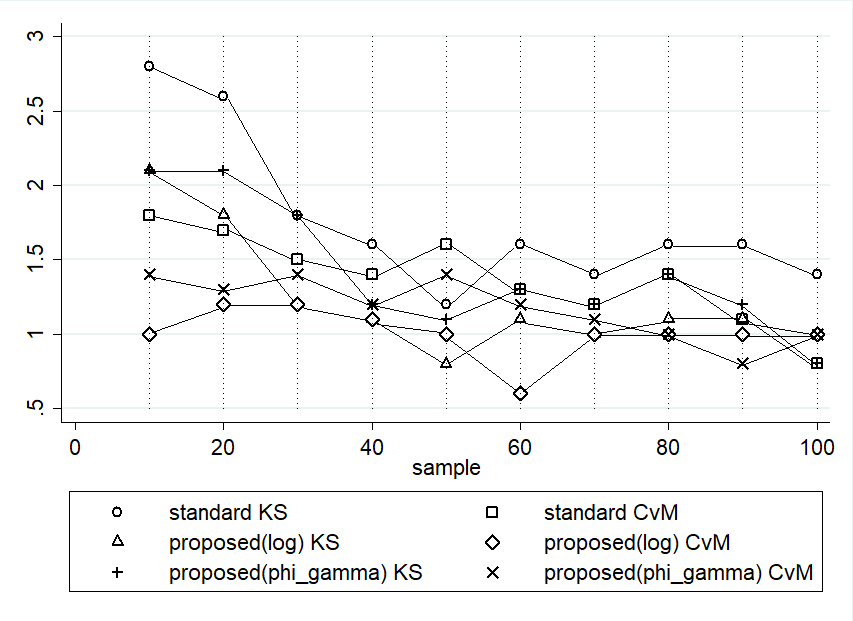}}}\\
	\subfigure[$H_0:\log.N(0,1)$\label{fig:weibull vs logn}]{
		\resizebox*{5cm}{!}{\includegraphics{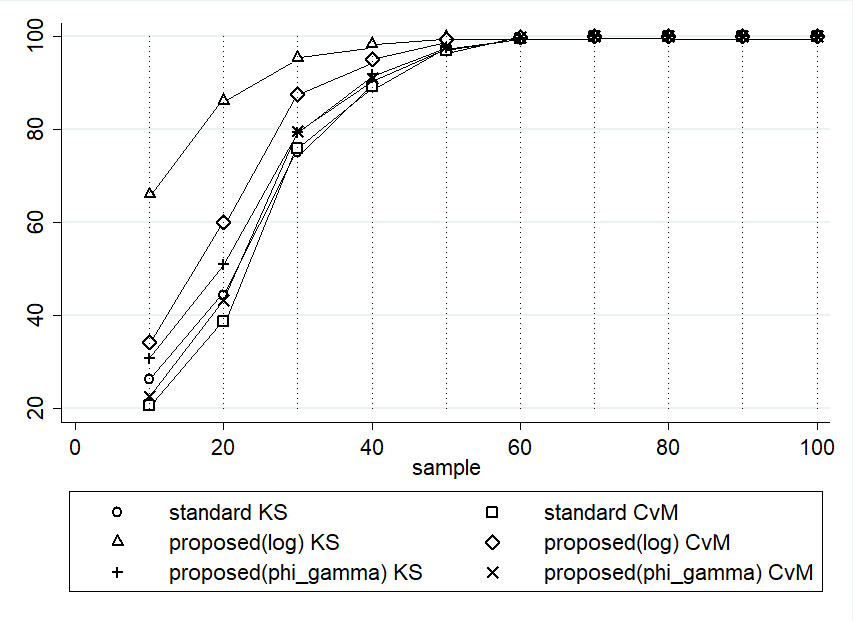}}}
	\subfigure[$H_0:abs.N(0,1)$\label{fig:weibull vs absn}]{
		\resizebox*{5cm}{!}{\includegraphics{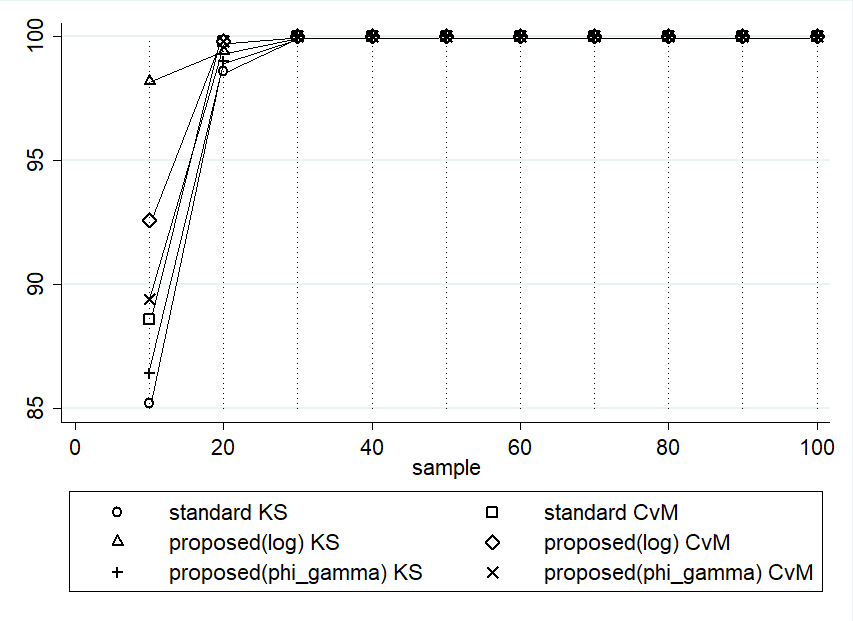}}}
	\caption{Simulated percentage (\%) of rejecting null hypothesis when the samples were drawn from $Weibull(2,2)$.}\label{fig:weibull tests}
\end{figure}

From Figure 3, we see that the modified KS and CvM tests outperformed the standard ones, especially the proposed KS test with logarithm as the bijective transformation. From Figure 3(a), (c), and (d), KS test with $\log$ function has the highest percentage of rejecting $H_0$ even when the sample sizes were still $10$. However, even though the new CvM test with logarithmic function was always the second highest in the beginning, $\widetilde{CvM}_{\log}$ was also the first one that reached $100\%$. On the other hand, based on Figure 3(b) we can say all statistical tests (standard and proposed) were having similar stable behaviours, as their numbers were still in the interval $0.5\%-2\%$. However at this time, $\widetilde{CvM}_{\log}$ performed slightly better than the others, because its numbers in general were the closest to $1\%$.

\begin{figure}
	\centering
	\subfigure[$H_0:Gamma(2,2)$\label{fig:logn vs gamma}]{
		\resizebox*{5cm}{!}{\includegraphics{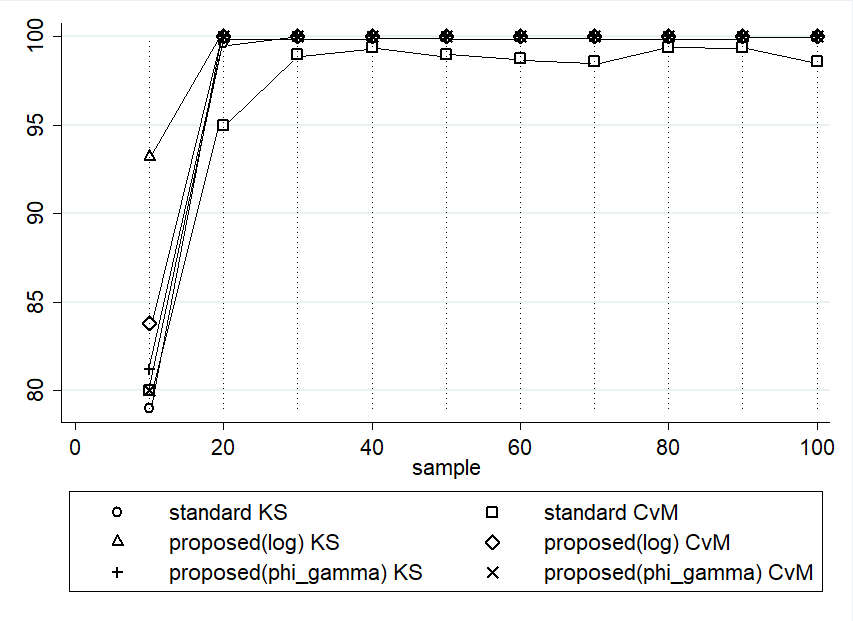}}}
	\subfigure[$H_0:Weibull(2,2)$\label{fig:logn vs weibull}]{
		\resizebox*{5cm}{!}{\includegraphics{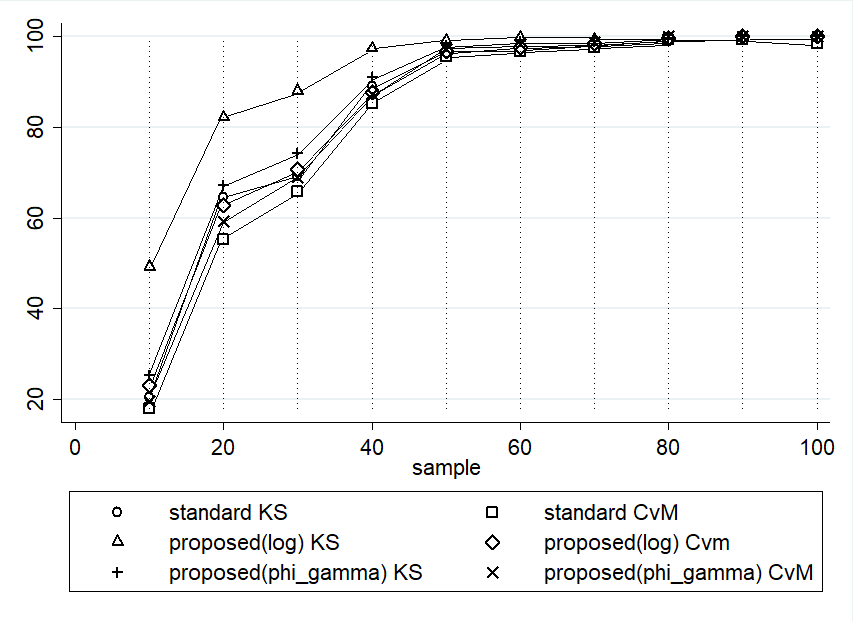}}}\\
	\subfigure[$H_0:\log.N(0,1)$\label{fig:logn vs logn}]{
		\resizebox*{5cm}{!}{\includegraphics{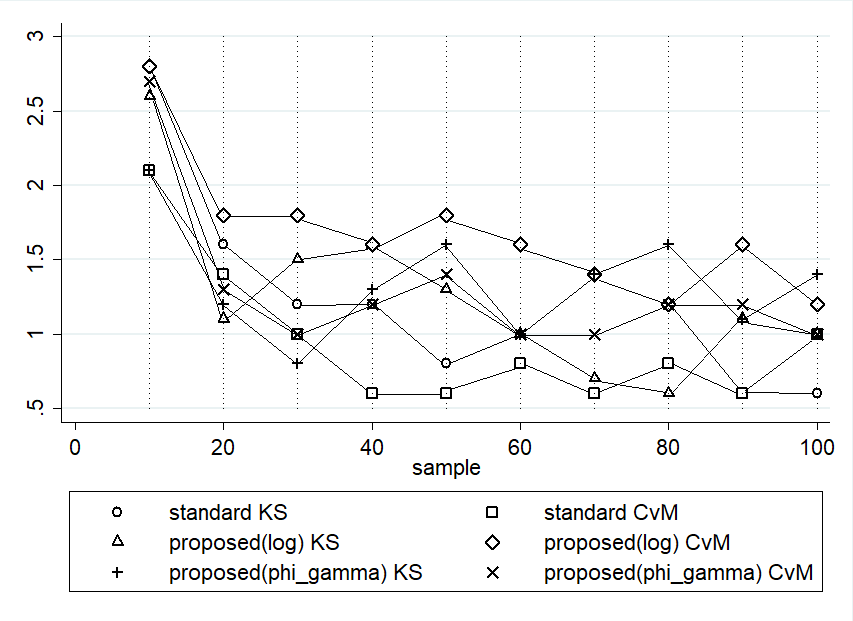}}}
	\subfigure[$H_0:abs.N(0,1)$\label{fig:logn vs absn}]{
		\resizebox*{5cm}{!}{\includegraphics{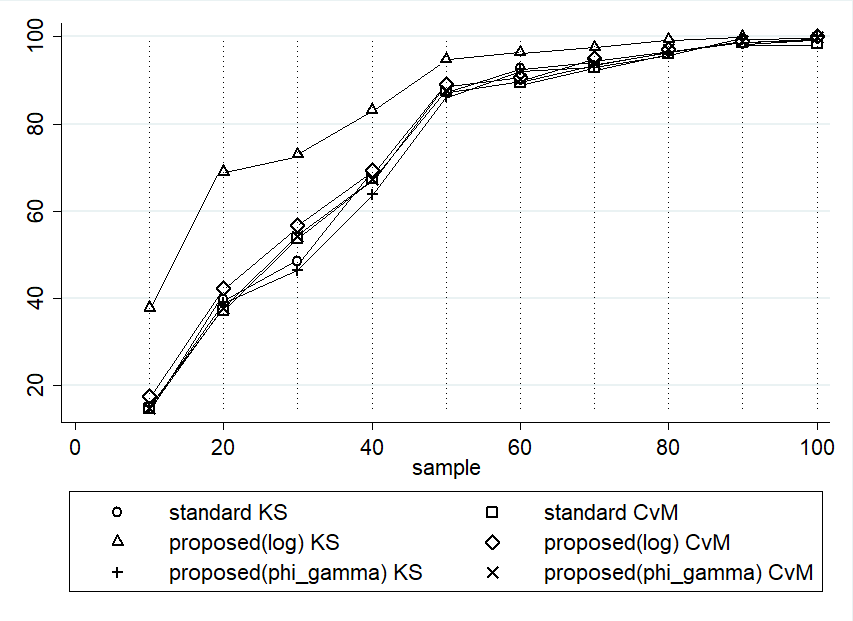}}}
	\caption{Simulated percentage (\%) of rejecting null hypothesis when the samples were drawn from $\log.N(0,1)$.}\label{fig:logn tests}
\end{figure}

Similar things happened when we drew the samples from the standard log-normal distribution, which our proposed methods outperformed the standard ones. However this time, the modified KS test with $g^{-1}=\log$ always gave the best results. Yet, we may take some notes from Figure 4. First, although when $n=10$ all the percentages were far from $1\%$ in Figure 4(c), but after $n=20$ every tests went stable inside $0.5\%-2\%$ interval. Second, as seen in Figure 4(d), it seems difficult to reject $H_0:abs.N(0,1)$ when the actual distribution is $\log.N(0,1)$, even $\widetilde{KS}_{\log}$ could only reach $100\%$ rejection after $n=80$. While, on the other hand, it was quite easy to reject $H_0:Gamma(2,2)$ as most of the tests already reached $100\%$ rejection when $n=20$ (similar to Figure 3(a)).

\begin{figure}
	\centering
	\subfigure[$H_0:U(0,1)$\label{fig:beta vs uniform}]{
		\resizebox*{5cm}{!}{\includegraphics{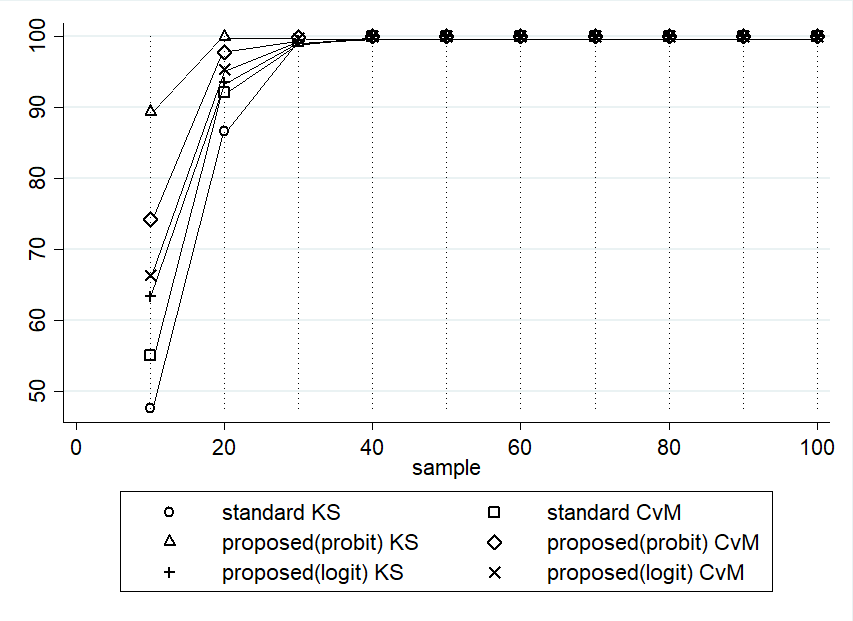}}}
	\subfigure[$H_0:Beta(1,3)$\label{fig:beta vs beta}]{
		\resizebox*{5cm}{!}{\includegraphics{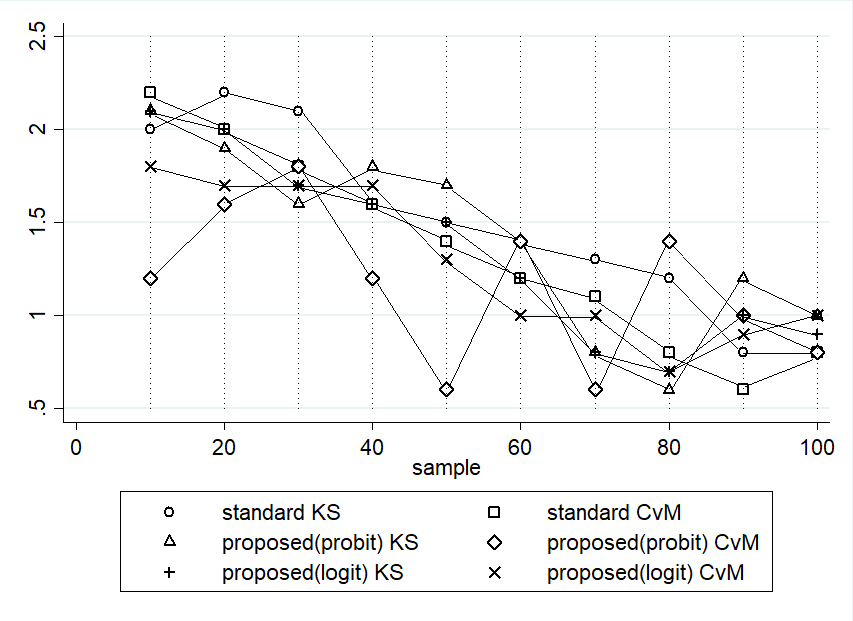}}}\\
	\subfigure[$H_0:Beta(2,2)$\label{fig:beta vs beta22}]{
		\resizebox*{5cm}{!}{\includegraphics{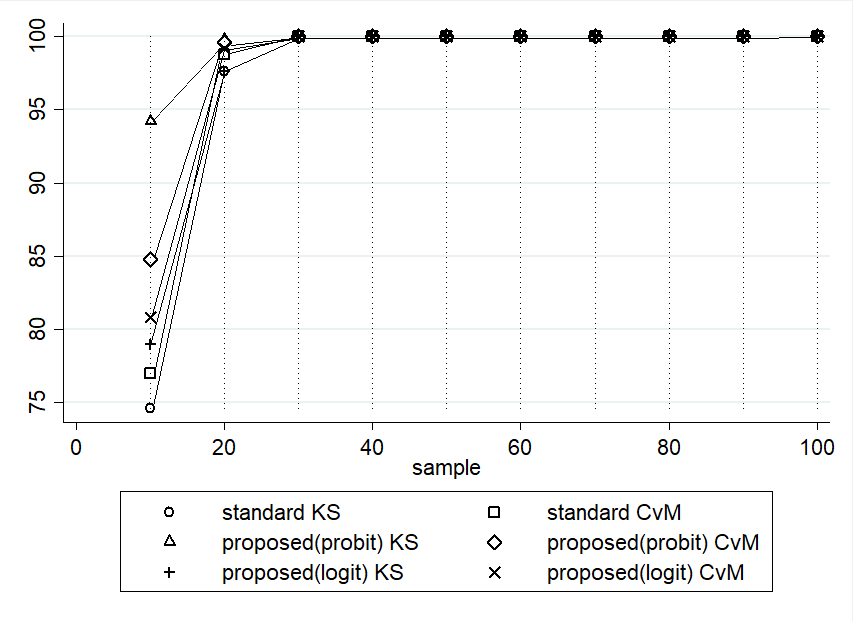}}}
	\subfigure[$H_0:Beta(3,1)$\label{fig:beta vs beta31}]{
		\resizebox*{5cm}{!}{\includegraphics{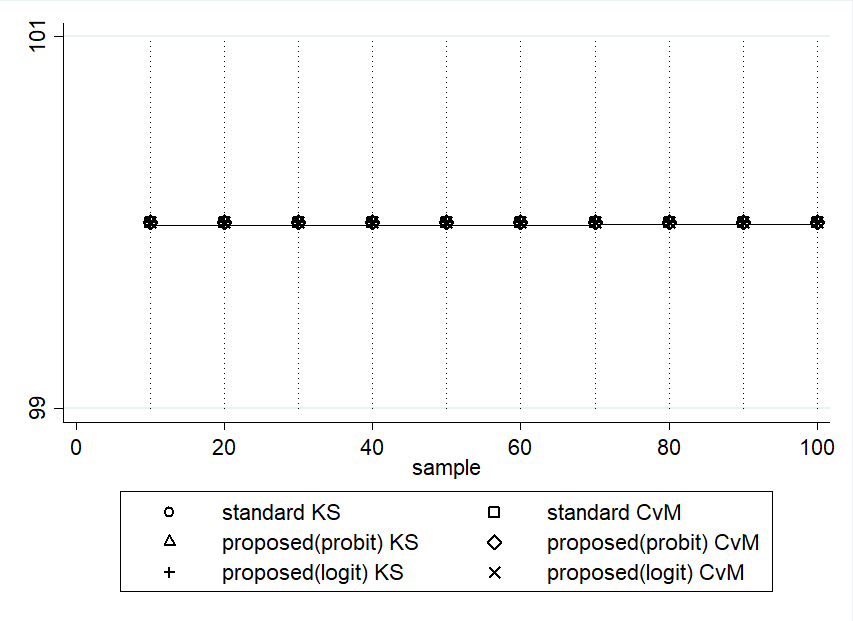}}}
	\caption{Simulated percentage (\%) of rejecting null hypothesis when the samples were drawn from $Beta(1,3)$.}\label{fig:beta tests}
\end{figure}

Something more extreme happened in Figure 5, as all of the tests could reach 100\% rejection rate since $n=30$, even since $n=10$ in Figure 5(d). Though seems strange, the cause of this phenomenon is obvious. The shape of the distribution function of $Beta(1,3)$ is so different with other three distributions in this study, especially with $Beta(3,1)$. Hence, even with small sample size, the tests could reject the false null hypothesis. However, we still are able to claim that our proposed tests worked better than the standard goodness-of-fit tests, because before all the tests reached 100\% point, the standard KS and CvM tests had the lowest percentages.

From this numerical studies, we can conclude that both the standard and the proposed KS and CvM tests will give the same result when the sample size is large. However, if the sample size is small, our proposed methods will give better and more reliable results.

\section{Conclusion}
This article has proposed new boundary-free kernel-smoothed Kolmogorov-Smirnov and Cram\'{e}r-von Mises tests when the data is supported on a proper subset of the real line. First we constructed a new boundary-free estimator of the distribution function using bijective transformations (we also intriduced a new boundary-free density estimator). After deriving the formulas of the bias and the variance of the estimator, we defined the new goodness-of-fit tests. The properties of our proposed methods have been discovered and discussed. Moreover, the results of the numerical studies reveal superior performances of the proposed methods.


\section*{Bibliography}
Abdous, B. 1993. Note on the minimum mean integrated squared error of kernel estimates of a distribution function and its derivatives. \textit{Communications in Statistics -- Theory and Methods} 22: 603--609. doi: 10.1080/03610929308831040

Antoneli, F. \textit{et al}. 2018. A Kolmogorov-Smirnov test for the molecular clock based on bayesian ensembles of phylogenies. \textit{PLoS ONE} 13(1): e0190826. doi: 10.1371/journal.pone.0190826

Azzalini, A. 1981. A note on the estimation of a distribution function and quantiles by a kernel method. \textit{Biometrika} 68: 326--328. doi: 10.2307/2335836

Baringhaus, L. and N. Henze. 2017. Cram\'{e}r-von Mises distance: Probabilistic interpretation, confidence intervals, and neighbourhood-of-model validation. \textit{Journal of Nonparametric Statistics} 29: 167--188. doi: 10.1080/10485252.2017.1285029

Chen, H., M. D\"{o}ring, and U. Jensen. 2018. Test for model selection using Cram\'{e}r-von Mises distance in a fixed design regression setting. \textit{AStA Advances in Statistical Analysis} 102: 505--535. doi: 10.1007/s10182-017-0317-0

Cowling, A., and P. Hall. 1996. On pseudodata methods for removing boundary effects in kernel density estimation. \textit{Journal of the Royal Statistical Society B} 58: 551--563. doi: 10.2307/2345893

Curry, J., X. Dang, and H. Sang. 2019. A rank-based Cram\'{e}r-von-Mises-type test for two samples. \textit{Brazilian Journal of Probability and Statistics} 33(3): 425--454. doi: 10.1214/18-BJPS396

Evans. D. L., J. H. Drew, and L. M. Leemis. 2017. The distribution of the Kolmogorov-Smirnov, Cramer-von Mises, and Anderson-Darling test statistics for exponential populations with estimated parameters. \textit{Computational Probability Applications} 247: 165--190. doi: 10.1007/978-3-319-43317-2\_13

Falk, M. 1983. Relative efficiency and deficiency of kernel type estimators of smooth distribution functions. \textit{Statistica Neerlandica} 37: 73--83. doi: 10.1111/j.1467-9574.1983.tb00802.x

Finner, H., and V. Gontscharuk. 2018. Two-sample Kolmogorov-Smirnov-type tests revisited: Old and new tests in terms of local levels. \textit{Annals of Statistics} 46(6A): 3014--3037. doi:10.1214/17-AOS1647

Hall, P., and T. E. Wehrly. 1991. A geometrical method for removing edge effects from kernel-type nonparametric regression estimators. \textit{Journal of the American Statistical Association} 86: 665--672. doi: 10.1080/01621459.1991.10475092

Hill, P. D. 1985. Kernel estimation of a distribution function. \textit{Communications in Statistics -- Theory and Methods} 14: 605--620. doi: 10.1080/03610928508828937

Jones, M. C., and P. J. Foster. 1996. A simple nonnegative boundary correction method for kernel density estimation. \textit{Statistica Sinica} 6: 1005--1013

Loeve, M (1963) \textit{Probability Theory}. New Jersey: Van Nostrand-Reinhold.

M\"{u}ller, H. G. 1991. Smooth optimum kernel estimators near endpoints. \textit{Biometrika} 78: 521--530. doi: 10.1093/biomet/78.3.521 

M\"{u}ller, H. G. 1993. On the boundary kernel method for nonparametric curve estimation near endpoints. \textit{Scandinavian Journal of Statistics} 20: 313--328

M\"{u}ller, H. G., and J. L. Wang. 1994. Hazard rate estimation under random censoring with varying kernels and bandwidths. \textit{Biometrics} 50: 61--76. doi: 10.2307/2533197

Nadaraya, E. A. 1964. Some new estimates for distribution functions. \textit{Theory of Probability and Its Applications} 9: 497--500. doi: 10.1137/1109069

Omelka, M., I. Gijbels, and N. Veraverbeke. 2009. Improved kernel estimation of copulas: weak convergence and goodness-of-fit testing. \textit{Annals of Statistics} 37: 3023--3058. doi: 10.1214/08-AOS666

Reiss, R. D. 1981. Nonparametric estimation of smooth distribution functions. \textit{Scandinavian Journal of Statistics} 8(2): 116--119

Schuster, E. F. 1985. Incorporating support constraints into nonparametric estimators of densities. \textit{Communication in Statistics -- Theory and Methods} 14: 1123-–1136. doi: 10.1080/03610928508828965

Shirahata, S., and I. S. Chu. 1992. Integrated squared error of kernel type estimator of distribution function. \textit{Annals of the Institute of Statistical Mathematics} 44: 579--591. doi: 10.1007/BF00050707

Singh, R. S., T. Gasser, and B. Prasad. 1983. Nonparametric estimates of distribution functions. \textit{Communications in Statistics -- Theory and Methods} 12: 2095--2108. doi: 10.1080/03610928308828593

Swanepoel, J. W. H. 1988. Mean integrated squared error properties and optimal kernels when estimating a distribution function. \textit{Communications in Statistics -- Theory and Methods} 17: 3785--3799. doi: 10.1080/03610928808829835

Watson, G. S., and M. R. Leadbetter. 1964. Hazard analysis 2. \textit{Sankhy\={a}: The Indian Journal of Statistics, Series A} 26: 101--106

Winter, B. B. 1973. Strong uniform consistency of integrals of density estimators. \textit{Canadian Journal of Statistics} 1: 247--253. doi: 10.2307/3315003

Yamato, H. 1973. Uniform convergence of an estimator of a distribution function. \textit{Bulletin of Mathematical Statistics} 15: 69--78

Yukich, J. E. 1989. A note on limit theorems for perturbed empirical processes. \textit{Stochastic Process and Applications} 33: 163--173. doi: 10.1016/0304-4149(89)90073-2

Zierk, J. \textit{et al}. 2020. Reference interval estimation from mixed distributions using truncation points and the Kolmogorov-Smirnov distance (kosmic). \textit{Scientific Reports} 10: 1704. doi: 10.1038/s41598-020-58749-2

\appendix

\section*{Appendix}

\subsection*{proof of theorem \ref{thm:biasvar}}
Utilizing the usual reasoning of \textit{i.i.d.} random variables and the transformation property of expectation, with $Y=g^{-1}(X_1)$, we have
\begin{eqnarray*}
	E[\widetilde{F}_X(x)]&=&\int_{-\infty}^\infty W\left(\frac{g^{-1}(x)-y}{h}\right)f_Y(y)\mathrm{d}y\\
	&=&\frac{1}{h}\int_{-\infty}^\infty F_Y(y)K\left(\frac{g^{-1}(x)-y}{h}\right)\mathrm{d}y\\
	&=&\int_{-\infty}^\infty F_Y(g^{-1}(x)-hv)K(v)\mathrm{d}v\\
	&=&\int_{-\infty}^\infty\left[F_Y(g^{-1}(x))-hv f_Y(g^{-1}(x))+\frac{h^2}{2}v^2 f_Y'(g^{-1}(x))+o(h^2)\right]K(v)\mathrm{d}v\\
	&=&F_X(x)+\frac{h^2}{2}c_1(x)\int_{-\infty}^\infty v^2 K(v)\mathrm{d}v+o(h^2),
\end{eqnarray*}
and we obtained the $Bias[\widetilde{F}_X(x)]$. For the variance of $\widetilde{F}_X(x)$, we first calculate
\begin{eqnarray*}
	E\left[W^2\left(\frac{g^{-1}(x)-g^{-1}(X_1)}{h}\right)\right]&=&\frac{2}{h}\int_{-\infty}^\infty F_Y(y)W\left(\frac{g^{-1}(x)-y}{h}\right)K\left(\frac{g^{-1}(x)-y}{h}\right)\mathrm{d}y\\
	&=&2\int_{-\infty}^\infty[F_Y(g^{-1}(x))-hvf_Y(g^{-1}(x))+o(h)]W(v)K(v)\mathrm{d}v\\
	&=&F_X(x)-2hg'(g^{-1}(x))f_X(x)r_1+o(h),
\end{eqnarray*}
and we got the variance.

\subsection*{proof of theorem \ref{thm:normal}}
For some $\delta>0$, using H\"{o}lder and Cram\'{e}r $c_r$ inequalities, we have
\begin{gather*}
E\left[\left|W\left(\frac{g^{-1}(x)-g^{-1}(X_1)}{h}\right)-E\left\{W\left(\frac{g^{-1}(x)-g^{-1}(X_1)}{h}\right)\right\}\right|^{2+\delta}\right]\\
\leq 2^{2+\delta}E\left[\left|W\left(\frac{g^{-1}(x)-g^{-1}(X_1)}{h}\right)\right|^{2+\delta}\right].
\end{gather*}
But, since $0\leq W(v)\leq 1$ for any $v\in\mathbb{R}$, then
\begin{gather*}
E\left[\left|W\left(\frac{g^{-1}(x)-g^{-1}(X_1)}{h}\right)-E\left\{W\left(\frac{g^{-1}(x)-g^{-1}(X_1)}{h}\right)\right\}\right|^{2+\delta}\right]\leq 2^{2+\delta}<\infty.
\end{gather*}
Also, because $Var\left[W\left(\frac{g^{-1}(x)-g^{-1}(X_1)}{h}\right)\right]=O(1)$, we get
\begin{gather*}
\frac{E\left[\left|W\left(\frac{g^{-1}(x)-g^{-1}(X_1)}{h}\right)-E\left\{W\left(\frac{g^{-1}(x)-g^{-1}(X_1)}{h}\right)\right\}\right|^{2+\delta}\right]}{n^{\delta/2}\left[Var\left\{W\left(\frac{g^{-1}(x)-g^{-1}(X_1)}{h}\right)\right\}\right]^{1+\delta/2}}\rightarrow 0
\end{gather*}
when $n\rightarrow\infty$. Hence, by Loeve (1963), and with the fact $\widetilde{F}_X(x)\rightarrow_p F_X(x)$, we can conclude its asymptotic normality.

\subsection*{proof of theorem \ref{thm:konsisten}}
Let $F_Y$ and $\widehat{F}_Y$ be the distribution function and the naive kernel distribution function estimator, respectively, of $Y_1,Y_2,...,Y_n$, where $Y_i=g^{-1}(X_i)$. Since $\widehat{F}_Y$ is a naive kernel distribution function, then Nadaraya (1964) guarantees that $\sup_{y\in\mathbb{R}}|\widehat{F}_Y(y)-F_Y(y)|\rightarrow_{a.s.}0$, which implies that
\begin{gather*}
\sup_{x\in\Omega}|\widehat{F}_Y(g^{-1}(x))-F_Y(g^{-1}(x))|\rightarrow_{a.s.}0.
\end{gather*}
However, because $F_Y(g^{-1}(x))=F_X(x)$, and it is clear that $\widehat{F}_Y(g^{-1}(x))=\widetilde{F}_X(x)$,
then this theorem is proven.

\subsection*{proof of theorem \ref{thm:biasvarpdf}}
Using the similar reasoning as in the proof of Theorem \ref{thm:biasvar}, we have
\begin{eqnarray*}
	E[\widehat{f}_X(x)]&=&\frac{1}{hg'(g^{-1}(x))}\int_{-\infty}^\infty K\left(\frac{g^{-1}(x)-y}{h}\right)f_Y(y)\mathrm{d}y\\
	&=&\frac{1}{g'(g^{-1}(x))}\int_{-\infty}^\infty f_Y(g^{-1}(x)-hv)K(v)\mathrm{d}v\\
	&=&\frac{f_Y(g^{-1}(x))}{g'(g^{-1}(x))}+\frac{h^2 f_Y''(g^{-1}(x))}{2g'(g^{-1}(x))}\int_{-\infty}^\infty v^2 K(v)\mathrm{d}v+o(h^2),
\end{eqnarray*}
and we obtained the bias formula. For the variance, first we have to calculate
\begin{eqnarray*}
	\frac{1}{hg'(g^{-1}(x))}E\left[K^2\left(\frac{g^{-1}(x)-Y}{h}\right)\right]&=&\frac{1}{g'(g^{-1}(x))}\int_{-\infty}^\infty f_Y(g^{-1}(x)-hv)K^2(v)\mathrm{d}v\\
	&=&f_X(x)\int_{-\infty}^\infty K^2(v)\mathrm{d}v+o(1),
\end{eqnarray*}
and the rests are easily done.

\subsection*{proof of theorem \ref{thm:ksequivalent}}
First, we need to consider the following inequality
\begin{eqnarray*}
	|KS_n-\widetilde{KS}|&=&\left|\sup_{v\in\Omega}|F_n(v)-F(v)|-\sup_{z\in\Omega}|\widetilde{F}_X(z)-F(z)|\right|\\
	&\leq&\sup_{x\in\Omega}\left||F_n(x)-F(x)|-|\widetilde{F}_X(x)-F(x)|\right|\\
	&\leq&\sup_{x\in\Omega}|F_n(x)-F(x)-\widetilde{F}_X(x)+F(x)|\\
	&=&\sup_{x\in\Omega}|\widetilde{F}_X(x)-F_n(x)|.
\end{eqnarray*}
Now, let $F_{n,Y}$ and $\widehat{F}_Y$ be the empirical distribution function and the naive kernel distribution function estimator, respectively, of $Y_1,Y_2,...,Y_n$, where $Y_i=g^{-1}(X_i)$. Hence, Omelka \textit{et al}. (2009) guarantees that $\sup_{y\in\mathbb{R}}|\widehat{F}_Y(y)-F_{n,Y}(y)|=o_p(n^{-1/2})$, which further implies that
\begin{gather*}
\sup_{x\in\Omega}|\widehat{F}_Y(g^{-1}(x))-F_{n,Y}(g^{-1}(x))|\rightarrow_p 0
\end{gather*}
with rate $n^{-1/2}$. But, $\widehat{F}_Y(g^{-1}(x))=\widetilde{F}_X(x)$ and $F_{n,Y}(g^{-1}(x))=F_n(x)$, and the equivalency is proven.

\subsection*{proof of theorem \ref{thm:cvmequivalnent}}
In this proof, we assume the bandwidth $h=o(n^{-1/4})$. Let us define
\begin{gather*}
\Delta_n=n\int_{-\infty}^\infty[\widetilde{F}_X(x)-F(x)]^2\mathrm{d}F(x)-n\int_{-\infty}^\infty[F_n(x)-F(x)]^2\mathrm{d}F(x).
\end{gather*}
Then, we have
\begin{eqnarray*}
	\Delta_n&=&n\int_{-\infty}^\infty\left[\widetilde{F}_X(x)-F(x)-F_n(x)+F(x)\right]\left[\widetilde{F}_X(x)-F(x)+F_n(x)-F(x)\right]\mathrm{d}F(x)\\
	&=&n\int_{-\infty}^\infty\frac{1}{n}\sum_{i=1}^n\left[W_i^*(x)-I_i^*(x)\right]\frac{1}{n}\sum_{j=1}^n\left[W_j^*(x)+I_j^*(x)\right]\mathrm{d}F(x),
\end{eqnarray*}
where
\begin{gather*}
W_i^*(x)=W\left(\frac{g^{-1}(x)-g^{-1}(X_i)}{h}\right)-F(x) \; \; \; \mathrm{and} \; \; \; I_i^*(x)=I(X_i\leq x)-F(x).
\end{gather*}
Note that if $i\neq j$, $W_i(\cdot)$ and $W_j(\cdot)$, also $I_i^*(\cdot)$ and $I_j^*(\cdot)$, are independent.

It follows from the Chauchy-Schwarz Inequality that
\begin{eqnarray*}
	E(|\Delta_n|)\leq n\int_{-\infty}^\infty\sqrt{E\left[\left\{\frac{1}{n}\sum_{i=1}^n(W_i^*(x)-I_i^*(x))\right\}^2\right]E\left[\left\{\frac{1}{n}\sum_{j=1}^n(W_j^*(x)+I_j^*(x))\right\}^2\right]}\mathrm{d}F(x).
\end{eqnarray*}
Let us define the bias
\begin{gather*}
b_n(x)=E\left[W\left(\frac{g^{-1}(x)-g^{-1}(X_i)}{h}\right)\right]-F(x)=O(h^2).
\end{gather*}
Hence, it follows from the independence that
\begin{eqnarray*}
	E\left[\left\{\frac{1}{n}\sum_{i=1}^n(W_i^*(x)-I_i^*(x))\right\}^2\right]&=&E\left[\left\{\frac{1}{n}\sum_{i=1}^n(W_i^*(x)-b_n(x)-I_i^*(x))\right\}^2\right]+b_n^2(x)\\
	&=&\frac{1}{n}E[\{W_1^*(x)-b_n(x)+I_1^*(x)\}^2]+b_n^2(x).
\end{eqnarray*}
Furthermore, we have
\begin{eqnarray*}
	E[\{W_1^*(x)-b_n(x)-I_1^*(x)\}^2]&=&E[\{W_1^*(x)-I_1^*(x)\}^2]-2b_n(x)E[W_1^*(x)-I_1^*(x)]+b_n^2(x)\\
	&=&E[\{W_1^*(x)\}^2-2W_1^*(x)I_1^*(x)-\{I_1^*(x)\}^2]-b_n^2(x).
\end{eqnarray*}
It follows from the mean squared error of $\widetilde{F}_X(x)$ that
\begin{gather*}
E[\{W_1^*(x)\}^2]=F(x)[1-F(x)]-2hr_1g'(g^{-1}(x))f_X(x)+O(h^2).
\end{gather*}
From the definition, we have
\begin{eqnarray*}
	E[W_1^*(x)I_1^*(x)]&=&E\bigg[W\left(\frac{g^{-1}(x)-g^{-1}(X_1)}{h}\right)I(X_1\leq x)-F(x)W\left(\frac{g^{-1}(x)-g^{-1}(X_1)}{h}\right)\\
	&& \; \; \; \; \; \; \; -F(x)I(X_1\leq x)+F^2(x)\bigg]\\
	&=&E\left[W\left(\frac{g^{-1}(x)-g^{-1}(X_1)}{h}\right)I(X_1\leq x)\right]-F^2(x)-b_n(x)F(x).
\end{eqnarray*}
For the first term we have
\begin{eqnarray*}
	E\left[W\left(\frac{g^{-1}(x)-Y}{h}\right)I(Y\leq g^{-1}(x))\right]&=&\int_{-\infty}^{g^{-1}(x)}W\left(\frac{g^{-1}(x)-y}{h}\right)f_Y(y)\mathrm{d}y\\
	&=&W(0)F_Y(g^{-1}(x))+\int_0^\infty F_Y(g^{-1}(x)-hv)K(v)dv\\
	&=&W(0)F(x)+F(x)\int_0^\infty K(v)dv+O(h).
\end{eqnarray*}
Since $K(\cdot)$ is symmetric around the origin, we have $W(0)=1/2$. Thus, we get
\begin{gather*}
E[W_1^*(x)I_1^*(x)]=F(x)[1-F(x)]+O(h).
\end{gather*}

Next we will evaluate $E\left[\{n^{-1}\sum_{i=1}^n(W_i^*(x)+I_i^*(x))\}^2\right]$. Using the bias term $b_n(x)$, we have
\begin{eqnarray*}
	E\left[\left\{\frac{1}{n}\sum_{i=1}^n(W_i^*(x)+I_i^*(x))\right\}^2\right]&=&E\left[\left\{\frac{1}{n}\sum_{i=1}^n(W_i^*(x)-b_n(x)+I_i^*(x))\right\}^2\right]+b_n^2(x)\\
	&=&\frac{1}{n}E[\{W_1^*(x)-b_n(x)+I_1^*(x)\}^2]+b_n^2.
\end{eqnarray*}
Based on previous calculations, we get
\begin{gather*}
E\left[\left\{\frac{1}{n}\sum_{i=1}^n(W_i^*(x)+I_i^*(x))\right\}^2\right]=O\left(\frac{1}{n}+h^4\right).
\end{gather*}
Therefore, if $h=o(n^{-1/4})$, we have $E(|\Delta_n|)=o(1)$. Using the Markov Inequality, we can show that $\Delta_n\rightarrow_p 0$, and then two statistics are equivalent under $H_0$.
\end{document}